\documentclass[aps,prl,twocolumn,showpacs,superscriptaddress]{revtex4}
\usepackage{epsfig}

\def\be{\begin{equation}} \def\ee{\end{equation}}
\def\bea{\begin{eqnarray}} \def\eea{\end{eqnarray}}

\def\Tau{\mathcal{T}}

\begin{document}

\title{Topological field theory and thermal responses of interacting topological superconductors}
\author{Zhong Wang}
\affiliation{Department of Physics, Stanford University,Stanford, CA
94305-4045}
\affiliation{Department of Modern Physics, University of Science and Technology of China, Hefei,  230026, China}
\affiliation{Institute for Advanced Study, Tsinghua University, Beijing, 100084, China,}
\author{Xiao-Liang Qi}
\affiliation{Department of Physics, Stanford University,Stanford, CA
94305-4045}
\author{Shou-Cheng Zhang}
\affiliation{Department of Physics, Stanford University,Stanford, CA
94305-4045}

\begin{abstract}
We investigate the three-dimensional, time-reversal invariant topological superconductors with generic interaction by their response to external fields. The first description is a gravitational topological field theory, which gives a $Z_2$ classification of topological superconductors, and predicts a half-quantized thermal Hall effect on the surface. The second description introduces an $s$-wave proximity pairing field on the surface, and the associated topological defects give an integer $Z$ classification of the topological superconductors.

\end{abstract}
 \maketitle

\section{Introduction}

Recently, topological insulators(TI) and topological superconductors(TSC) have attracted great attention in
condensed matter physics\cite{qi2010a,hasan2010,qi2010b,moore2010}. Topological insulators are states that are incompressible in the bulk but have topological surface states protected by the bulk topological properties which remain robustly metallic upon arbitrary perturbation on the surface, within the given symmetry class. Time-reversal invariant topological insulators in two and three dimensions (2D and 3D) have been theoretically predicted and experimentally realized in recent years\cite{bernevig2006c,koenig2007,fu2007a,hsieh2008,zhang2009,
xia2009,chen2009}. Topological superconductors are analog of topological insulators in superconductors, which have a full superconducting gap, and gapless Majorana edge states propagating on the boundary. Time-reversal breaking topological superconductors in 2D\cite{read2000} and time-reversal invariant topological superconductors in 2D and 3D\cite{qi2009b,schnyder2008,roy2009a} have been theoretically predicted but have not been verified experimentally. The most promising candidate for 3D topological superconductor (more precisely, superfluid) is the $^3$He-B phase.

There are two approaches for the classification and description of TIs and TSCs:  topological field theory(TFT) and topological band theory(TBT). In the TFT approach, topological insulators are defined by the quantized responses derived from low-energy effective actions\cite{qi2008}. TFT also reveals the relations among TIs in various dimensions. For instance, the 4D Chern-Simons insulator\cite{Zhang2001} is the root state of 3D and 2D time reversal invariant(TRI) TIs. The 3D topological insulator is characterized by the topological effective theory $S=\frac{\theta}{32\pi^2} F_{\mu\nu}F_{\sigma\tau}\epsilon^{\mu\nu\sigma\tau}$ with $\theta=\pi$ mod $2\pi$, which describes the electromagnetic response of the topological insulator when time-reversal symmetry is broken on the surface.\cite{qi2008,qi2009} The TBT was first developed in the work of Kane and Mele\cite{kane2005a,kane2005b}
on the $Z_2$ topological classification of quantum spin Hall insulator, which was later generalized to 3D TRI TIs\cite{fu2007b, Roy2009,moore2007} and more recently to all ten symmetry classes\cite{kitaev2009,schnyder2008}. It was also shown that TFT reduces to TBT in the non-interacting limit for time-reversal invariant TIs\cite{wang2010a}.

Interestingly, in 1D it has been shown that the non-interacting classifications are sometimes unstable to interaction.\cite{fidkowski2010}. Therefore, in order to define TSCs in dimensions higher than 1, it is helpful to apply the topological field theory (TFT) approach of TIs\cite{Zhang2001,qi2008}, since the TFT describes physical response properties of the system, which is applicable to interacting systems. All topological insulators can be described in this way by TFT\cite{qi2008}, including the 2D and 3D TIs. An explicit expression of the topological invariant of these classes in interacting systems are obtained in the form of Green's function\cite{wang2010b}. However, there are some symmetry classes in which this approach does not apply. For example, the 3D time-reversal invariant TSC is one of them. Due to the absence of charge conservation symmetry, no TFT can be obtained by the coupling to the electromagnetic field.

In this paper, we introduce two new approaches to the classification of 3D interacting time-reversal invariant TSCs. The general idea behind that can generate a physical observable topological response of the system. Such topological response can then be used to classify the topological states. TIs and TSCs in different topological classes and different dimensions require different external fields to probe their topological response properties. We present two different approaches with different probe fields used, which applly to different physical settings. The first approach is the gravitational TFT, which describes the topological response of the TSC to a gravitational field. Since the coupling to gravitational
field does not require charge conservation, the gravitational TFT can be applied to TSC similar to the electromagnetic TFT approach of 3D TI\cite{qi2008}. The topological term is the gravitational analogy of the $\theta$-term for the gauge field, which is the Pontryagin invariant of the spacetime manifold\cite{Nakahara1990}.
However, similar to the electromagnetic $\theta$ term for 3D TI, the gravitational TFT only provides a $Z_2$ classification of the 3D TSC. In the non-interacting limit, the nontrivial class with $\theta=\pi$ corresponds to the TSC states with odd topological quantum number, while the trivial class with $\theta=0$ corresponds to those with even topological quantum number.

A natural question is whether this indicates that with interaction only a $Z_2$ classification is stable, or there are other ways we can define the $Z$ classification in interacting TSC. To answer this question we propose the second approach. We consider the proximity effect of the surface of 3D TSC with an s-wave superconductor. The proximity effect introduce an s-wave pairing field to the surface states. When the pairing field has a $\pi$ phase domain wall on the surface, there are $N$ number of chiral Majorana fermions propagating on the domain wall. Since the chiral Majorana fermions are stable without any symmetry requirement, and can be characterized by the thermal current it carries, one can use the number of Majorana zero mode as a generic definition of TSC in 3D. From this approach we conclude that the integer classification of 3D TSC remains robust when interaction is considered. This approach also directly leads to an experimental proposal for the measurement of the topological quantum number. Moreover, the idea of defining interacting TIs and TSCs by studying the defects obtained from some symmetry-breaking field configuration can be generalized to generic symmetry classes.



\section{Gravitational topological field theory of 3D TSC}
First we consider gravitational TFT of time-reversal invariant (TRI) TSC in 3D. For simplicity, we first investigate continuous models using the example of the $^3$He-B phase as a
3D TRI TSC.

In the flat space-time, the Hamiltonian of $^3$He-B phase is given by\cite{qi2009b,chung2009}
\bea
H({\bf p})=m({\bf p})\tau^3 + \Delta {\bf \sigma}\cdot {\bf p} \tau^1 \label{HofHe3B}
\eea
in the Nambu space $\psi = (c_\uparrow,c_\downarrow,c^\dagger_\downarrow,-c^\dagger_\uparrow)$, where $\uparrow,\downarrow$ denote spin orientation. $\sigma^{1,2,3}$ and $\tau^{1,2,3}$ labels the Pauli matrices in spin and particle-hole spaces, respectively. $m({\bf p})=p^2/2M-\mu$ is the kinetic energy term. The weak pairing phase with $M>0, \mu>0$ is topologically nontrivial and corresponds to the physical $^3$He-B phase\cite{qi2009b,schnyder2008,roy2009}, while the strong pairing phase with $\mu<0$ is a trivial superfluid phase. After rescaling the momenta $\Delta p_i\rightarrow p_i$, the corresponding Lagrangian can be written as \bea \textit{L}=\overline{\psi}[p_\mu\Gamma^\mu+m(p_i)]\psi \eea where $\mu=0,1,2,3$(We use $(0,1,2,3)$ interchangeably with $(t,x,y,z)$), $\Gamma^0=\tau^3$ and $\Gamma^i=i\sigma^i\tau^2;i=1,2,3$. This is exactly the Majorana-Dirac Lagrangian if $m(p_i)=m$ is independent of $p_i$.

Now we turn to Lagrangian of $^3$He-B in curved space-time.  It is convenient to take the space-time as a closed 4-manifolds $M$. To consistently define the Lagrangian globally,  $M$ is required to be a spin manifold\cite{lawson1989}. Locally the low energy action is \bea S_\psi = \int_M d^4x \sqrt{-g} \overline{\psi}[i\Gamma^\alpha e_\alpha^\mu(\partial_\mu+\frac{1}{2}i\omega_\mu^{\beta\gamma}\Sigma_{\beta\gamma})+m]\psi \eea where $\Sigma_{\alpha\beta}\equiv \frac{1}{4}i[\Gamma^\alpha,\Gamma^\beta]$ is the generator of Lorentz transformation in the spinor representation, and $e_\alpha^\mu$ is the vielbeins\cite{Nakahara1990}. Physically we can interpret the gravitational field as an external source coupling to the energy-momentum tensor of the fluid. In a deeply insightful paper, Volovik points out that the
order parameter of the $^3$He-B phase also couples to fermions like a gravitational vielbein\cite{volovik1990}. Therefore, we can also interpret $e_\alpha^\mu$ as the internal order parameter of the $^3$He-B phase.

Because the Majorana fermions are gapped, we can integrate out them to obtain a gravitational effective action. The term we are interested in is the gravitational theta term, which can be easily obtained by calculating chiral anomaly due to gravity. Due to the non-invariance of fermion Jacobian, a chiral transformation $\psi \rightarrow \exp(i\Gamma^5\theta/2)\psi$, which inverts $m$ to $-m$, generates the
topological term in the effective action\cite{fujikawa1979, eguchi1976} as
$ S_\theta =  \frac{\theta}{2} \int_M  \hat{A}(M)$, where $\hat{A}(M)$ is the Dirac genus of $M$\cite{Nakahara1990}.
Since the Hamiltonian with positive mass $m=-\mu>0$ is adiabatically connected to a trivial superconductor by taking $\mu\rightarrow -\infty$, by a chiral rotation of $\theta=\pi$ we can obtain the action of the $^3$He-B phase with $\mu>0$ as $S_\theta =  \frac{\pi}{2} \int_M  \hat{A}(M)$. It is worth noting that Majorana fermion has a $1/2$ coefficient compared to Dirac fermion. More explicitly, we have the Pontryagin invariant
\bea
S_\theta=-\frac{\theta}{1536\pi^2}\int d^4 x \epsilon^{\mu\nu\rho\sigma}R^\alpha_{\beta\mu\nu}R^\beta_{\alpha\rho\sigma}\label{GravityTFT}
\eea
with $\theta=\pi$.  For a generic $4$-manifold, $S_\theta=n\theta/48$ with integer $n$. However, for spin manifolds the value of Dirac genus is restricted to $\int_M \hat{A}(M)=2n$ with integer $n$\cite{atiyah1959}, so that $S_\theta=n\theta$, as is expected from time reversal invariance. On a domain wall between $\theta=0$ and $\theta=\pi$, we get the Chern-Simons term from the theta term as
\bea S_{2D}= \frac{1}{384\pi}\int d^3x\epsilon^{\mu\nu\rho}{\rm Tr}(\omega_\mu \partial_\nu \omega_\rho + \frac{2}{3}\omega_\mu \omega_\nu \omega_\rho ) \label{CS} \eea
which describes the surface of $^3$He-B phase in the presence of surface magnetization.


For a generic TSC with a non-relativistic Hamiltonian, the direct coupling to gravity may be complicated. However, physically one can always consider the non-relativistic Hamiltonian as a low energy limit of an underlying relativistic theory, so that the coupling to gravity is in principle always well-defined, and the topological term (\ref{GravityTFT}) has a coefficient of $\theta=0$ or $\pi$ mod $2\pi$ as is required by TR symmetry. The topological term (\ref{GravityTFT}) is insensitive to an arbitrary deformation of the fermion Hamiltonian as long as the fermions remain gapped and TRI is preserved. For a non-interacting TSC with topological number $N$, its Hamiltonian can always be deformed to $N$ copies of the $^3$He-B Hamiltonian\cite{kitaev2009}, so that the topological term has $\theta=N\pi$. 
However, for even integer $N$, $\exp(iS_\theta)=1$ and $S_\theta$ has no physical effect. We conclude from this fact that the gravitational TFT gives a $Z_2$ classification of 3D TRI TSC, which is weaker than the integer classification of the non-interacting system. Because gravitational responses are related to thermal responses\cite{luttinger1964}, the gravitational TFT predicts topological thermal responses, as we shall discuss in the next section.

\section{Surface state description} To understand the difference between $Z_2$ and $Z$ classification, it is helpful to study the topological surface states. For the $^3$He-B phase, the surface state is a single flavor of 2D massless Majorana fermion\cite{qi2009b,chung2009}.  For our discussion here, it is most convenient to use the Majorana basis in which the surface state effective Hamiltonian is given by
\begin{eqnarray}
H=\sum_{\bf k}\eta_{-\bf k}^T v\left(\sigma_zk_x+\sigma_xk_y\right)\eta_{\bf k}
\end{eqnarray}
with $\eta_{\bf k}=\eta_{-\bf k}^\dagger$ the two-component Majorana fermion field. The two components carry opposite spin, and the time-revesal operation is defined $T^{-1}\eta_{\bf k}T=i\sigma_y\eta_{-\bf k}$. It should be noticed that only Pauli matrices $\sigma_z$ and $\sigma_x$ are allowed in this Hamiltonian since the single particle Hamiltonian $h({\bf k})=v\left(\sigma_zk_x+\sigma_xk_y\right)$ satisfies the antisymmetry condition $h(-{\bf k})=-h^T({\bf k})$. No mass term is allowed by time-reversal symmetry since the only possible mass term $i\sigma_y$ breaks time-reversal symmetry. More generically for a topological superconductor with topological quantum number $N$, the surface state consists of multiple copies of the Majorana fermions with different chirality:
\begin{eqnarray}
H&=&\sum_{s=1}^{N_+}\sum_{\bf k}\eta_{s,-\bf k}^T v\left(\sigma_zk_x+\sigma_xk_y\right)\eta_{s,\bf k}\nonumber\\
& &+\sum_{s=1}^{N_-}\sum_{\bf k}\psi_{s,-\bf k}^T v\left(\sigma_zk_x-\sigma_xk_y\right)\psi_{s,\bf k}\label{genericsurfacetheory}
\end{eqnarray}
where $\eta_{s,{\bf k}}$ and $\psi_{s,{\bf k}}$ are the ``left-handed" and ``right-handed" Majorana fermions, and the integer $N_+-N_-=N$ is determined by the bulk topological invariant. Such a definition of ``chirality" for Majorana fermion is possible because $\sigma_y$ is not allowed to appear in the linear-$k$ terms, therefore the two kinds of Majorana fermions $\eta_{\bf k}$ and $\psi_{\bf k}$ cannot be deformed to each other.

More rigorously, the topological invariant $N=N_+-N_-$ can be defined by the following formula:
\begin{eqnarray}
N=-\sum_i\frac12{\rm Ind}\left.\left(\mathcal{T}\left[\frac{\partial h({\bf k})}{\partial k_x},\frac{\partial h({\bf k})}{\partial k_y}\right]\right|_{{\bf k}={\bf K}_i}\right)
\end{eqnarray}
with $\mathcal{T}=i\sigma_y{\rm I}$ the time-reversal transformation
matrix for all Majorana fermions, and ${\bf K}_i$ runs over all time-reversal invariant momenta $(0,0),(0,\pi),(\pi,0),(\pi,\pi)$ in the surface Brillouin zone. ${\rm Ind}(M)$ for a Hermitian
matrix $M$ is defined as the number of positive eigenvalues minus the
number of negative eigenvalues.

From the discussion above we have seen from the surface state point of view why the topological classification is $Z$ for non-interacting theory. The relation of this surface state picture to the gravitational TFT (\ref{GravityTFT}) is similar to that between the Dirac surface state and the topological electromagnetic action of 3D TI\cite{qi2008}. Consider a generic T-breaking mass term of the surface states $H_m=\sum_sm_s\eta_{s-\bf k}^T\sigma_y\eta_{s\bf k}+\sum_sm'_s\psi_{s-\bf k}^T\sigma_y\psi_{s\bf k}$. With the mass terms the surface states are gapped and we obtain a quantized thermal Hall coefficient
\bea \kappa_{xy}= \frac{\pi k_B^2 T}{24\hbar}\left(\sum_{s=1}^{N_+} {\rm sgn}(m_s)-\sum_{s=1}^{N_-} {\rm sgn}(m'_s)\right) \label{thermalHall}\eea
Each Majorana cone contributes half the thermal Hall coefficient of that of a 2D $(p+ip)$ superconductor\cite{blote1986,affleck1986,kane1997,read2000}. Such a thermal Hall effect on the surface is consistent with the description of the bulk TFT (\ref{GravityTFT}) since the Pontryagin invariant can be reduced to a gravitational Chern-Simons term on the boundary as eq.(\ref{CS}). However, due to the dependence on the sign of mass terms, the thermal Hall coefficient is only determined by the surface state theory (\ref{genericsurfacetheory}) mod $2\times \frac{\pi k_B^2 T}{24\hbar}$. This is consistent with the observation that the coefficient of the bulk TFT (\ref{GravityTFT}) takes $Z_2$ value $0$ or $\pi$ mod $2\pi$. In other words, only the topological superconductors with odd $N$ necessarily have thermal Hall effect on the surface, while those with even $N$ can have zero thermal Hall coefficient {\it for some choice of mass terms}. It's interesting to note that the same gravitational TFT approach can
be applied to time-reversal invariant TI, for which the surface
thermal Hall conductivity will be doubled. Thus we see that the
gravitational TFT description is trivial for 3D TI.

\section{Integer $Z$ classification of 3D TRI superconductor by topological defects}
The analysis above seems to indicate that only the topological superconductors with odd topological quantum number are stable. However, in the following we will show that the $Z$ classification is actually robust with interaction, although it cannot be captured by the gravitational TFT. As has been discussed in the introduction part, a general philosophy of classifying interacting topological phases is by defining a physical topological response of the system to a proper external field. The gravitational TFT can only probe the $Z_2$ part of the topological invariant, but the full $Z$ can be probed by another physical field---an s-wave pairing field. We consider the generic 3D time-reversal invariant Majorana fermion system with a Hamiltonian $H=\int d^3{\bf r}d^3{\bf r'}\eta^T({\bf r})h({\bf r},{\bf r'})\eta({\bf r'})+H_{\rm int}$. Here $h$ is an anti-symmetric Hermitian single-particle operator, and $H_{\rm int}$ describes non-quadratic terms in Majorana fermion $\eta$. The Hamiltonian is time-reversal invariant satisfying $T^{-1}HT=H$, with the anti-unitary time-reversal operator defined by $T^{-1}\eta({\bf r}) T=\mathcal{\Tau}\eta({\bf r})$ and $\Tau^2=-1$. Now we introduce the following perturbation term to the Hamiltonian
\bea
H_\Delta&=&-i\int d^3{\bf r}\Delta({\bf r})\eta^T({\bf r})\left[T^{-1}\eta({\bf r}) T\right]\nonumber\\
&=&\int d^3{\bf r}i\Delta({\bf r})\eta^T({\bf r})\mathcal{\Tau}\eta({\bf r})\label{swavefield}
\eea
with $\Delta({\bf r})$ being a real field. To see the physical meaning of this term, we can take the $^3$He-B Hamiltonian (\ref{HofHe3B}) as an example. In the complex fermion basis $H_\Delta=-\int d^3{\bf r}i\Delta({\bf r})\left(c_{\uparrow\bf k}^\dagger({\bf r}) c_{\downarrow{\bf k}}^\dagger({\bf r})-h.c.\right)$ which is an $s$-wave pairing term with imaginary pairing order parameter. It is important to notice that such a T-breaking pairing term (\ref{swavefield}) is completely determined by the representation of time-reversal transformation and is independent of the basis choice and any other detail of the system.

Now we consider the time-reversal invariant TSC with the perturbation (\ref{swavefield}) only near the surface. Physically, such a field can be induced by the proximity effect between the TSC and an s-wave superconductor on top of it. Since this term breaks time-reversal symmetry, the surface states of the TSC can become gapped. The effect of this $s$-wave pairing field to the surface state can be seen straightforwardly from the surface effective theory (\ref{genericsurfacetheory}). Since $T^{-1}\eta_sT=i\sigma_y\eta_s,~T^{-1}\psi_tT=i\sigma_y\psi_t$ for the surface Majorana fermions, the $s$-wave pairing term (\ref{swavefield}) is reduced to $H_{\Delta{\rm surf}}=\int d^2{\bf r}\Delta({\bf r})\left(\sum_s\eta^T_s\sigma_y\eta_s+\sum_t\psi_t^T\sigma_y\psi_t\right)$ on the surface. In other words, an uniform pairing field $\Delta({\bf r})=\Delta>0$ corresponds to a uniform mass term $m_s=m_t'=\Delta$ for all the surface Majorana fermions. Consequently the thermal Hall conductivity given by Eq. (\ref{thermalHall}) is $\kappa_{xy}=N\frac{\pi k_B^2 T}{24\hbar}$ with $N$ the bulk topological invariant, so that by introducing the pairing field, the topological quantum number $N$ can be obtained without ambiguity.

\begin{figure}
\includegraphics[width=7.0cm, height=3.5cm]{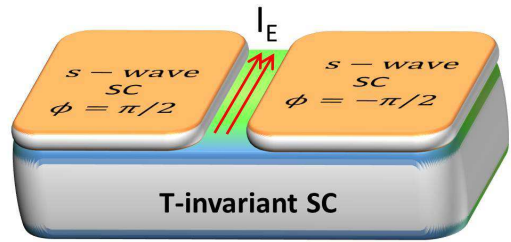}
\caption{Illustration of the experimental setting for the measurement of the integer-valued topological quantum number of 3D TSC. $s$-wave pairing is introduced on the surface by proximity effect, with a $\pi$-Josephson junction. The topological quantum number $N$ of the bulk is characterized by the number of chiral Majorana fermions propagating along the junction illustrated by the red arrows, which carry a quantized thermal current $I^E=N\frac{\pi k_B^2T^2}{24\hbar}$ for low temperature $T\rightarrow 0$.}
\label{fig:swavedomain}
\end{figure}

The statement above can be made more explicit by considering a domain wall on the surface where $\Delta({\bf r})$ changes sign, as shown in Fig. \ref{fig:swavedomain}. Physically, such a domain wall is a $\pi$-Josephson junction of the s-wave superconductor on top of the time-reversal invariant TSC. For the surface Majorana fermions, the mass term changes sign across the domain wall, therefore there are chiral Majorana fermion propagating along the domain wall\cite{read2000}. On comparison, similar $\pi$-Josephson junction on the surface of topological insulator leads to non-chiral Majorana fermions\cite{fu2008}.  Due to the uniform mass for all Majorana fermions, the number of chiral Majorana fermion is given by $N_+$ while the number of anti-chiral Majorana fermion is given by $N_-$. Consequently there are $N=N_+-N_-$ copies of chiral Majorana fermions which are robust gapless states on the domain wall.  It is important to notice that a chiral Majorana fermion in 1D is well-defined and robust even in an interacting system, since it has a chiral central charge $c=1/2$ and thus carries a chiral thermal current $I^E=\frac{\pi k_B^2T^2}{24\hbar}$. Thus one can use the chiral thermal current as a physical observable way to define chiral Majorana fermions, and the chiral Majorana fermions on the domain wall can be used as a generic definition of the bulk 3D TSC as follows:
\begin{itemize}
\item For a generic 3D time-reversal invariant superconductor, apply a small pairing field in the form of Eq. (\ref{swavefield}) with a domain wall of $\Delta({\bf r})$, as shown in Fig. \ref{fig:swavedomain}. The superconductor is a TSC with topological quantum number $N\in{\rm Z}$ if the thermal current propagating along the domain wall at temperature $T$ is $I^E=N\frac{\pi k_B^2T^2}{24\hbar}$ in the $T\rightarrow 0$ limit.
\end{itemize}

We thank E Berg, T. Hughes, S. Ryu, S.Raghu, J. Teo, Y.S. Wu and S.L. Wan for helpful discussions.  Upon finishing this work we became aware of a paper by S. Ryu, J. E. Moore and A. W. W. Ludwig\cite{ryu2010b} which proposed that the $Z$ classification of 3D TSC can be characterized by the gravitational response. However, as discussed in this paper, we conclude that the gravitational TFT can only characterize the $Z_2$ part of the classification, while the $Z$ classification has to be defined by introducing the $s$-wave pairing field. This work was supported by the NSF under grant numbers DMR-0904264 and the US Department of Energy, Office of Basic Energy Sciences under contract DE-AC02-76SF00515. ZW acknowledges the support of CSC and NSF of China(Grant No.10675108). XLQ acknowledges the support of the Sloan Foundation.

\bibliography{gravity,TI}

\end{document}